\documentclass[aps,pre,preprint,groupedaddress,showpacs] {revtex4-1}
\usepackage[dvips]{graphicx}
\usepackage{textcomp}
\usepackage{amssymb}
\usepackage{amsmath}
\usepackage{dsdshorthand}

\begin{document}

\title{
\Large\bf Two- and three-point functions at criticality: \\
 Monte Carlo simulations of the three-dimensional $(q+1)$-state clock model
}

\author{Martin Hasenbusch}

\affiliation{
Institut f\"ur Theoretische Physik, Universit\"at Heidelberg,
Philosophenweg 19, 69120 Heidelberg, Germany}

\begin{abstract}
We simulate the improved $(q+1)$-state  clock model on the simple cubic lattice 
at the critical point on lattices of a linear size up to $L=960$.
We compute operator product expansion (OPE) coefficients for the 
three-dimensional XY universality class. These are compared with 
highly accurate estimates obtained by using the conformal bootstrap method.
We find that the results are consistent. 
\end{abstract}

\maketitle
\section{Introduction}
In recent years substantial progress in critical phenomena in three 
dimensions has been achieved by using the conformal bootstrap (CB) method.
For reviews, see, for example, \cite{Simmons-Duffin:2016gjk,PaRyVi18}.
In particular, in the case of the three-dimensional Ising universality class,
the results for critical exponents are considerably more accurate than 
those obtained by other methods \cite{Kos:2016ysd,Simmons-Duffin:2016wlq}.
Very recently also accurate results for the three-dimensional 
XY universality class were provided \cite{che19}. 

In addition to critical exponents, the CB provides accurate estimates for
so called operator product expansion (OPE) coefficients $\lambda_{ijk}$. 
These are defined by the behavior of three-point functions at the critical
point.
The OPE coefficients are difficult to access by other methods. In the case 
of the three-dimensional Ising universality class, only recently results 
have been obtained by using Monte Carlo simulations of lattice models
 \cite{Ca15,Co16,He17,myStructure}. These are far less precise than
those obtained by using the CB. However, the agreement of the 
results from the lattice and CB gives further support for the fact that 
both methods examine the same renormalization group (RG) fixed point.

The functional form of two-point 
functions of primary operators is fixed by conformal invariance
\begin{equation}
\label{twopoint}
 \langle  \cO_1(x_1)  \cO_2(x_2)  \rangle =
  \frac{C_1 \delta_{\Delta_1,\Delta_2}}{|x_1 -  x_2|^{2 \Delta_1}} \;\;,
\end{equation}
where $\cO_i$ is the operator taken at the site $x_i$, and
$\Delta_i$ is its scaling dimension. 

Also the form of three-point functions is fixed by conformal invariance.
Normalizing the operators such that $C_i=1$, eq.~(\ref{twopoint}),
one gets \cite{Polyakov}
\begin{equation}
\label{threepoint}
\langle  \cO_1(x_1)  \cO_2(x_2)   \cO_3(x_3) \rangle
= \frac{\lambda_{123}}{
       |x_1-x_2|^{\Delta_1+\Delta_2-\Delta_3}
       |x_2-x_3|^{\Delta_2+\Delta_3-\Delta_1}
       |x_3-x_1|^{\Delta_3+\Delta_1-\Delta_2}
                } \;\;,
\end{equation}
where the OPE coefficients $\lambda_{123}$  depend on the universality class.
For a detailed discussion see for example 
the lecture note \cite{Simmons-Duffin:2016gjk}.

In the present work, we apply the idea of ref. \cite{myStructure} to the 
XY universality class in three dimensions.
To this end, we simulate  the improved $O(2)$-symmetric $\phi^4$ model 
and the improved
$(q+1)$-state clock model on the simple cubic lattice at the critical 
temperature. 
To reduce the statistical error of the two and three-point 
functions, we use a variance reduction method \cite{Pa83,LuWe01}.  To 
reduce finite size effects, large linear lattice sizes $L$ are considered.
In our simulations, we go up to $L=960$. On top of that, an extrapolation
to $L \rightarrow \infty$ is still needed.  
Our estimates for the OPE coefficients
turn out to be consistent with those obtained by using the CB.

In table \ref{CFTdata} we summarize results for the scaling dimensions 
$\Delta_i$ and the OPE coefficients $\lambda_{ijk}$ obtained by using the CB.
In the case of the scaling dimensions we give the most accurate results
obtained from Monte Carlo simulations \cite{Xu19,myClock,HaVi11} of lattice
models for comparison.  In the 
case of $\Delta_s$,  we also give the estimate obtained by analyzing 
specific heat data for $^4$He near the $\lambda$-transition 
\cite{Lipa96,Lipa00,Lipa03}.  Note that the scaling dimensions are 
related with the critical exponents that are usually discussed in critical
phenomena \cite{PeVi}. In particular the critical exponent of the correlation 
length is given by $\nu=1/(3-\Delta_s)$ and the exponent of the 
correlation function at criticality $\eta= 2 \Delta_{\phi} -1$.
The estimate of $\nu$ obtained by using high temperature (HT) series and
Monte Carlo simulations of lattice models \cite{XY1,XY2} differs from that
obtained from experiments \cite{Lipa96,Lipa00,Lipa03} by several times 
the combined error. Recent Monte Carlo studies \cite{Xu19,myClock} and the CB
work \cite{che19} confirm the results of refs. \cite{XY1,XY2}. 

\begin{table}
\caption{\sl \label{CFTdata} Scaling dimensions and OPE coefficients for 
the three-dimensional XY universality class.
Comparison of conformal bootstrap (CB) results \cite{che19,Kos:2016ysd}
with estimates from Monte Carlo (MC) or experiment (EXP). The leading
charge 0, 1, and 2 scalars are denoted by s, $\phi$, t, respectively.
For a discussion of the meaning of the errors that are quoted see the 
references.
}
\begin{center}
\begin{tabular}{ccclll}
\hline
Quantity & method & value & ref. \\
\hline 
$\Delta_s$ &  EXP  & 1.50946(22) & \cite{Lipa96,Lipa00,Lipa03} \\
           &  MC   & 1.51153(40) & \cite{Xu19} \\
           &  MC   & 1.51122(15) & \cite{myClock} \\
           &  CB   & 1.51136(22) & \cite{che19} \\
           &  CB   & 1.5117(25)  & \cite{Kos:2016ysd} \\
\hline 
$\Delta_{\phi}$ & MC   &  0.51927(24)  & \cite{Xu19} \\
                & MC   &  0.519050(40) & \cite{myClock} \\
                & CB   &  0.519088(22) & \cite{che19} \\
                & CB   &  0.51926(32)  & \cite{Kos:2016ysd} \\
\hline
$\Delta_t$ &      MC   & 1.2361(11)   & \cite{HaVi11} \\
           &      CB   &  1.23629(11) & \cite{che19} \\
\hline 
$\lambda_{\phi \phi s} $ & CB  & 0.687126($27$) & \cite{che19} \\
                         & CB  & 0.68726(65)      & \cite{Kos:2016ysd} \\
$\lambda_{s s s} $       & CB  & 0.830914($32$) & \cite{che19} \\
                         & CB  & 0.8286(60)       & \cite{Kos:2016ysd} \\
$\lambda_{t t s} $       & CB  & 1.25213($14$)  & \cite{che19} \\
$\lambda_{\phi \phi t} $ & CB  & 1.213408($65$) & \cite{che19} \\
\hline
\end{tabular}
\end{center}
\end{table}

The outline of the paper is the following. In section \ref{theModel} we
define the models that are simulated and we summarize numerical results, 
for example for the critical temperature, which are used in our simulations.
Next, in section \ref{observables} 
we define the observables and briefly recall the variance 
reduction method. In section \ref{numerics} we discuss the simulations 
and analyze our numerical results. Finally, we conclude and give an outlook.

\section{The lattice models}
\label{theModel}
We performed preliminary simulations by using the $O(2)$-symmetric $\phi^4$ 
model on the lattice.
The final results were obtained from simulations of the $(q+1)$-state 
clock model with $q=32$.  Note that in the limit $q \rightarrow \infty$, the 
dynamically diluted XY model studied in refs. \cite{XY1,XY2} is reached.
Both models have a parameter that can be tuned such that leading corrections
to scaling vanish.  Models taken at a good approximation of this value
are denoted as improved. The idea to study improved models to
get better precision on universal quantities goes back to refs. 
\cite{ChFiNi,FiCh}. For a discussion, see, for example, section 2.3 of the 
review \cite{PeVi}. 

In the following, we define the 
models and summarize estimates of the improved models and   the inverse 
critical temperature given in the literature.

\subsection{The $O(2)$-symmetric $\phi^4$ model}
The $O(N)$-symmetric $\phi^4$ model on the simple cubic lattice is defined by 
the reduced Hamiltonian
\begin{equation}
 {\cal H}_{\phi^4}= -\beta \sum_{<xy>} \vec{\phi}_x \cdot  \vec{\phi}_y
+ \sum_x \left [ \vec{\phi}_x^{\,2} + \lambda (\vec{\phi}_x^{\,2} -1)^2  \right]
\;,
\end{equation}
where $\vec{\phi}_x \in \mathbb{R}^N$ is the field at the site 
$x=(x^{(0)},x^{(1)},x^{(2)})$, where $x^{(i)} \in \{0,1,2,...,L_i-1 \}$. 
Here we are labeling the components of $x$ by an upper index.
A lower index is used to discriminate different sites on the lattice.
$\left<xy\right>$ denotes a pair of nearest neighbor sites on the simple
cubic lattice. In our simulations $L_0=L_1=L_2=L$ throughout. In the present 
work we consider the case $N=2$. 

For the $O(2)$-symmetric $\phi^4$ model on the simple cubic lattice 
the authors of ref. \cite{XY2} find for the improved model 
$\lambda^* = 2.15(5)$ and $\beta_c=0.5091503(3)[3]$ and $0.5083355(3)[4]$ for
$\lambda=2.1$ and $2.2$, respectively.
These estimates are obtained by requiring that $(Z_a/Z_p)^*=0.3203(1)[3]$,  
where $Z_p$ and $Z_a$ are the partition functions for a system with periodic 
boundary conditions in all directions and anti-periodic in one direction and
periodic in the remaining ones, respectively. The superscript $^*$ refers to 
the fixed point value for the given lattice geometry. The number quoted in $()$
refers to the statistical error obtained in a specific fit, while the number 
given in $[]$ is an estimate of the systematic error.
In the case of $\beta_c$, 
the number given in $[]$ is the error due to the uncertainty of $(Z_a/Z_p)^*$.
Here we have reanalyzed 
unpublished data generated in 2013 for $\lambda=2.1$ using the estimates 
$(Z_a/Z_p)^*= 0.32037(6)$ and $(\xi_{2nd}/L)^*=0.59238(7)$ given in table 3 
of ref. \cite{myClock} as input. We arrive at
\begin{equation}
\label{betacphi2.1}
\beta_c(\lambda=2.1) = 0.5091504(1) \;\;, 
\end{equation}
where the number quoted in $()$ includes both the statistical as  well as 
the systematical error.
We simulate the $O(2)$-symmetric $\phi^4$ model by using a hybrid of local 
Metropolis, local overrelaxation and single cluster \cite{Wolff} updates.
For a discussion of this algorithm see for example Appendix A of 
ref. \cite{HaVi11}.

\subsection{The $(q+1)$-state clock model}
The model can be viewed as a generalization of the $q$-state clock model.
The field $\vec{s}_x$ at the site 
$x=(x^{(0)},x^{(1)},x^{(2)})$, where $x^{(i)} \in \{0,1,2,...,L_i-1 \}$,
might assume one of the following values
\begin{equation}
\vec{s}_x \in
\left\{(0,0), \left(\cos(2 \pi m/q), \sin(2 \pi m/q)    \right)
\right\} \;,
\end{equation}
where $m \in \{1,...,q\}$. In our simulations we take $L_0=L_1=L_2=L$ 
throughout. Compared with the $q$-state clock model,
$(0,0)$ is added as possible value of the field variable. In our simulation 
program,
we store the field variables by using labels $m =0,1,2,...,q$. We assign
\begin{equation}
 \vec{s}(0) = (0,0)
\end{equation}
and for $m>0$
\begin{equation}
 \vec{s}(m) =  \left(\cos(2 \pi m/q), \sin(2 \pi m/q)   \right) \;\;.
\end{equation}
The reduced Hamiltonian is given by
\begin{equation}
\label{ddXY}
 {\cal H} = -  \beta \sum_{\left<xy\right>}  \vec{s}_x \cdot
     \vec{s}_y -D  \sum_x \vec{s}_x^{\,2} - \vec{H} \sum_x \vec{s}_x \;.
\end{equation}
In our simulations, we consider a vanishing external field $\vec{H} = 
\vec{0}$ throughout. We introduce the weight factor
\begin{equation}
\label{weight}
w(\vec{s}_x)  = \delta_{0,\vec{s}_x^{\,2}} +  
 \frac{1}{q} \delta_{1,\vec{s}_x^{\,2}}
              = \delta_{0,m_x} + \frac{1}{q} \sum_{n=1}^q \delta_{n,m_x}
\end{equation}
that gives equal weight to $(0,0)$ and the collection of all values 
$|\vec{s}_x| =1$. Now the partition function can be written as
\begin{equation}
 Z = \sum_{\{\vec{s}\} }  \prod_x w(\vec{s}_x) \; \exp(-{\cal H}) \;,
\end{equation}
where $\{\vec{s}\}$ denotes a configuration of the field.

Note that in the limit $q \rightarrow \infty$, we recover the dynamically
diluted
XY (ddXY) model studied in refs. \cite{XY1,XY2}. The reduced Hamiltonian
of the ddXY model has the same form as eq.~(\ref{ddXY}):
\begin{equation}
\label{ddXY2}
{\cal H}_{ddXY} =  -  \beta \sum_{\left<xy\right>}  \vec{\phi}_x \cdot
\vec{\phi}_y - D  \sum_x \vec{\phi}_x^{\,2} - \vec{H} \sum_x \vec{\phi}_x \;,
\end{equation}
where $\vec{\phi}_x$ is a vector with two real components.
The partition function is given by
\begin{equation}
 Z = \prod_x \left[\int d\mu(\phi_x) \right] \; \exp(- {\cal H}_{ddXY})  \;,
\end{equation}
with the local measure
\begin{equation}
d\mu(\phi_x) =  d \phi_x^{(1)} \, d \phi_x^{(2)} \,
\left[
\delta(\phi_x^{(1)}) \, \delta(\phi_x^{(2)})
 + \frac{1}{2 \pi} \, \delta(1-|\vec{\phi}_x|)
\right] \; .
\label{lmeasure}
\end{equation}

In ref. \cite{myClock} we simulated the model with $q=8$. We find
$D^*= 1.058(13)$, see eq.~(63) of  \cite{myClock}. For nearby values 
of $D$ we obtain 
\begin{eqnarray}
\label{betac105q8}
 \beta_c(D=1.05) &=& 0.56082390(10) \;, \\
 \beta_c(D=1.07) &=& 0.55888340(10) \;. 
\end{eqnarray}
In the appendix B 2 of ref. \cite{myClock} we study the $q$-dependence 
of non-universal quantities such as the critical temperature. We find that
already for $q=8$ the estimates differ only slightly from those
for the limit $q \rightarrow \infty$.  At the level of our 
statistical accuracy, estimates for $q \ge 10$ can not be distinguished from
those for the limit $q \rightarrow \infty$.
Taking the results of the appendix B 2 of ref. \cite{myClock}, we arrive at 
\begin{eqnarray}
\label{betac105qL}
 \beta_c(D=1.05) &=& 0.56082418(10)[10] \;, \\
 \beta_c(D=1.07) &=& 0.55888368(10)[10] \;
\end{eqnarray}
for $q \ge 10$. The number in $[]$ gives the uncertainty of the 
extrapolation.  The major part of the simulations here is performed 
for $q=32$. In this case 6 bits are needed to store the field variable 
at one site. Also the arrays needed to store possible changes in the
weight that are used to speed up the Metropolis and cluster updates 
are still small enough to fit into the cache of the CPU.
We use a hybrid of local Metropolis and single cluster updates \cite{Wolff} to 
simulate the model. For a detailed discussion see section IV of ref. 
\cite{myClock}.

\section{The observables}
\label{observables}
Let us define the observables measured on the finite lattice.
Note that in our measurements, following
ref. \cite{myStructure} we replace the field at the site $x$ by the 
sum of its six nearest neighbors. The idea is that statistical noise is 
reduced and furthermore in the case of the $(q+1)$-state clock model the
rotational invariance is better approximated.

Let us define the correlation functions that are measured in the simulations.
To this end,
we use the notation of the $O(N)$-invariant $\phi^4$ model.
In the following we denote the components of the field variable $\vec{\phi}_x$
by $\phi_{x,i}$ with $i \in \{0,1, ...,N-1\}$.
We study correlation functions of $\phi$ and the two derived 
quantities $s$ and $t$. The observables are defined for $N \ge 2$. In 
our numerical study discussed below, we consider the case $N=2$. 
The scalar $s$ with charge $0$ is given by 
\begin{equation}
s_x = \sum_i \phi_{x,i} \phi_{x,i}  - \bar{s} \;,
\end{equation}
where $\bar{s} =\langle \sum_i \phi_{x,i} \phi_{x,i} \rangle$ for the 
given lattice size.
The scalar with charge $2$ is given by the traceless compound 
\begin{equation}
t_{x,ij}  = \phi_{x,i} \phi_{x,j} - \delta_{ij}  \frac{\phi_{x}^2}{N} \;.
\end{equation}
Note that these lattice quantities also contain scaling fields with 
the same symmetry properties but larger scaling dimensions than the lowest.
Furthermore, conformal invariance is only well approximated at length scales
considerably larger than the lattice spacing. 
Therefore, the correlation functions show corrections at small distances.
In the case of improved models, the leading correction should  be related 
with the breaking of the rotational invariance of the continuum by the 
lattice. The corresponding correction exponent is $\omega_r = 2.02(1)$
\cite{XY1,XY2,ROT98}.
For a recent discussion of corrections to scaling in an improved 
model see section III of ref. \cite{myClock}.

The two-point functions, without any normalization  are 
\begin{eqnarray}
\label{gpp}
  g_{\phi \phi}(x_1,x_2) &=& \sum_{i}  \langle \phi_{x_1,i} \phi_{x_2,i} \rangle
\;, \\
\label{gss}
  g_{s s}(x_1,x_2) &=&  \langle s_{x_1} s_{x_2} \rangle
\;, \\
\label{gtt}
  g_{t t}(x_1,x_2) &=& \sum_{ik} \langle t_{x_1,i,k} t_{x_2,i,k} \rangle \;.
\end{eqnarray}
We consider the following three-point correlation functions:  
\begin{eqnarray}
\label{Gpps}
  G_{\phi \phi s}(x_1,x_2,x_3) &=& 
\sum_{i} \langle \phi_{x_1,i} \phi_{x_2,i} s_{x_3} \rangle \;, \\
\label{Gsss}
  G_{s s s}(x_1,x_2,x_3) &=& 
\langle s_{x_1} s_{x_2} s_{x_3} \rangle \;, \\
\label{Gtts}
  G_{t t s}(x_1,x_2,x_3) &=&      
\sum_{ik} \langle t_{x_1,i,k} t_{x_2,i,k} s_{x_3} \rangle \;, \\
\label{Gppt}
  G_{\phi \phi t}(x_1,x_2,x_3) &=&      
\sum_{ik} \langle \phi_{x_1,i} \phi_{x_2,k} t_{x_3,i,k}  \rangle \;. 
\end{eqnarray}
Note that our normalizations of the two- and three-point functions are 
not the same as those of refs. \cite{che19,Kos:2016ysd}. This leads to
a factor of $\sqrt{2}$ in the result for $\lambda_{\phi \phi t}$, while
it cancels in the other three cases \cite{private}.

\subsection{Our choices for $x_1$, $x_2$, and $x_3$}
The variance reduction method requires that the lattice is subdivided 
into blocks. For technical reasons, we compute correlation functions 
only for the sites at the center of these blocks. These sites are given 
by $x^{(i)} = n_s k_i$.  In our simulations we have used the three different 
choices $n_s=2$, $4$, and $6$. Throughout the linear lattice size $L$ is a 
multiple of $n_s$ and $k_i \in \{0, 1, ...,L/n_s-1 \}$. In the following 
we refer to $n_s$ as stride.
In order to keep the study tractable, we have to single out
a few directions for the displacements between the points.
In the case of the two-point functions we consider displacements
along the axes, the face diagonals, and the space diagonals.
In the following, these are indicated by ($a$), ($f$), and ($d$), respectively.
In our simulation program we summed over all choices that are related by
symmetry to reduce the statistical error. 
In the following we shall denote the two-point
function by $g_{r,\cO_1,\cO_2}(x)$, where $r \in \{a,f,d\}$ gives the
direction and $x=|x_1-x_2|$ is the distance between the two points.
In the case of the three-point functions
\begin{equation}
G_{r,\cO_1,\cO_2,\cO_3}(x)=\langle \cO_1(x_1) \cO_2(x_2)
\cO_3(x_3) \rangle
\end{equation}
we consider two different geometries that are indicated by $r$.
For $r=f$ the largest displacement is along a face diagonal. For example
\begin{equation}
 x_3-x_1 = (j,0,0) \;\; , \; x_3-x_2 = (0,j,0) \;.
\end{equation}
Our second choice is indicated by $r=d$ and the largest displacement
is along a space diagonal.  For example
\begin{equation}
 x_3-x_1 = (j,0,0) \;\; , \; x_3-x_2 = (0,j,j) \;,
\end{equation}
where $j=n_s k$, where $k$ is integer. Also here we sum in our simulation
over all choices that are related by symmetry to reduce the statistical
error. The argument $x$ of $G$ gives the smallest distance between two points
$x=j$.  

In order to eliminate the constants $C_i$, eq.~(\ref{twopoint}), and the power
law behavior from the
three-point functions, we directly normalized our estimates of the
three-point functions by the corresponding ones of two-point functions.
For the direction $r=f$ we get for example
\begin{equation}
\label{ppt}
\lambda_{\phi \phi t} \simeq   2^{-\Delta_{t}/2 }
\frac{G_{f,\phi \phi t}(x)}
     {g_{a, \phi \phi}(x) \; g^{1/2}_{f, t t}(\sqrt{2} x)} \;.
\end{equation}
Based on the numbers given in table \ref{CFTdata} we used
as numerical values $\Delta_{t}= 1.23629$, $\Delta_{s}=1.5113$, and
$\Delta_{\phi}=0.51908$ for the scaling dimensions. 
Note that the systematical error of the estimate
of $\lambda_{ijk}$ due to the uncertainty of the scaling dimensions is 
negligible. 

\subsection{Variance reduced measurement}
The variance reduction method used here is based on the ideas of 
refs. \cite{Pa83,LuWe01}.  The method is discussed in detail in section
V of ref. \cite{myStructure}. Here we summarize the basics for completeness.

In the case of an $N$-point correlation function, the lattice is partitioned 
into $N$ areas $B_i$, where each of these areas contains
one of the sites $x_1$, $x_2$, ..., $x_N$.  These areas are chosen such 
that for each pair $i \ne j$ none of the sites in $B_i$ is a nearest 
neighbor of a site in $B_j$. Let us denote the collection of the remaining 
sites as $R$. Now the sampling of the correlation function can be 
reorganized in the following way.

In a straight forward approach one would estimate the expectation value
of the $N$-point correlation function by averaging over $M$ configurations
\begin{equation}
\overline{ \cO_1(x_1)  \cO_2(x_2) ... \cO_N(x_N) } = 
\frac{1}{M} \sum_{\alpha} 
\cO_{1,\alpha}(x_1)  \cO_{2,\alpha}(x_2) ... \cO_{N,\alpha}(x_N) \;,
\end{equation}
where $\alpha$ labels configurations that have been generated by using a Markov 
chain. We assume that the process is equilibrated and the configurations 
are generated with a probability density proportional to the Boltzmann
factor. $\cO_{i,\alpha}(x_i)$ denotes the value of $\cO_{i}(x_i)$ assumed for 
configuration $\alpha$. 

In the case of the variance reduced measurement, we first average 
$\cO_{i}(x_i)$ over configurations on $B_i$ that have been generated,  while
keeping the field on $R$ fixed:
\begin{equation}
\overline{ \cO_1(x_1)  \cO_2(x_2) ... \cO_N(x_N) } = 
\frac{1}{M} \sum_{\alpha} \overline{ \cO_{1,\alpha}(x_1)} \; 
\overline{\cO_{2,\alpha}(x_2)} ... 
\overline{\cO_{N,\alpha}(x_N)} \;,
\end{equation}
where 
\begin{equation}
 \overline{\cO_{i,\alpha}(x_i)} = \frac{1}{m} \sum_{\gamma} 
\cO_{i,\alpha,\gamma}(x_i) \;.
\end{equation}
Here we have generated $m$ configurations labeled by $\gamma$
 on $B_i$, keeping the 
field on $R$ fixed. The configurations on $R$ are labeled by $\alpha$. 
The effect of this averaging for each site separately is that we 
consider $m^N$ configurations for the $N$-point function.  For
small $m$ this translates into 
\begin{equation}
 \epsilon^2 \propto  \frac{1}{m^N}
\end{equation} 
for the statistical error $\epsilon$ of the estimate of the $N$-point 
correlation function. As $m$ increases, the effect of fixing the 
configuration on $R$ becomes visible and $\epsilon^2$ converges to 
a finite limit as $m \rightarrow \infty$ and can be reduced only by
increasing $M$.   
There is in general an optimal value of $m$. This value of $m$ depends
on the choice of $\cO_{i}$ and the distances. 
Finding a good choice of $m$ requires
some numerical experimentation.  Below we shall specify our implementation
of this general idea.

We only used the sites $(j_0 n_s,j_1 n_s,j_2 n_s)$, with
$j_i \in \{0,n_s, 2 n_s, ..., L_i/n_s-1\}$ for the measurements of the
two- and three-point functions.
As areas we consider blocks of the size $l_b^3$, where $l_b=2 n_s -1$. 
The sites used for the measurement are at the center of the blocks.

Computing the block averages we used local updates only. In particular 
in the case of the $(q+1)$-clock model, we used the first version of 
the Metropolis update discussed in section IV. A. of ref. \cite{myClock}. 
Computing averages for the blocks, keeping the remainder $R$ fixed, 
we update more frequently towards the center of the blocks.  To this 
end we perform a cycle of updates, similar to the cycle used in a 
multigrid updating scheme.  In particular, we sweep over subblocks
of the size $3^3$, $5^3$, ..., $l_b^3$. In addition, as smallest subset,
we consider the central site and its 6 nearest neighbors. For each of these
sweeps we perform a measurement. The frequency $n_x$ of the sweeps is 
chosen such that the number of sites times $n_x$ is roughly the same for 
all sizes. For example in the case $n_s=6$, where $l_b=11$, in one such 
cycle 268 measurements are performed. In our production runs for $n_s=6$, 
we performed $160$ update cycles for a given configuration on the remainder 
$R$. Hence in total $160 \times 268 = 42880$ measurements are performed 
for a given configuration on $R$.

Note that for our setup two blocks with the central sites $x_1$ and $x_2$ 
are separated if $|x_1^{(i)}-x_2^{(i)}| \ge 2 n_s$ for at least 
one direction $i$. Computing the two- and three-point functions, one therefore
has to note that only results for $|x_k^{(i)}-x_l^{(i)}| \ge 2 n_s$ for 
at least one direction $i$ are valid.

The simulation is built up in the following way: 
First we equilibrate the system without measuring
by performing 2000 times the following sequence of updates:
One sweep with the Metropolis update type two, one sweep with the Metropolis 
update type one, and $L$ times a single cluster update. These updates are 
discussed in  section IV of ref. \cite{myClock}.

For each measurement, we performed ten times the
following sequence of updates: two Metropolis sweeps followed by
$L$ single cluster updates.  In the first and the sixth sequence, the
first Metropolis is of type two, while all others are of type one.
Note that here the measurements, including the updates of the blocks for
variance reduction, are far more expensive than the updates of the system 
as a whole. Therefore between the measurements, a relatively large number 
of updates is performed, in order to measure on essentially uncorrelated 
configurations.  
In principle the final configurations of  separated blocks could be used
as update of the main Markov chain. In our case a $1/2^3$ of the blocks
could be used to this end. For simplicity we abstained from doing so.

\subsection{Finite size effects}
\label{fse}
Compared with the linear size $L$ of the lattice, the distances that
we consider for our two- and three-point functions are small. In that respect,
they can be viewed as local scalar operators with charge $0$ such as the
energy density.
The energy density on a finite lattice of the linear size $L$, 
for a vanishing external field, behaves  as
\begin{equation}
\label{finiteL}
 E(\beta_c,L) = c L^{-\Delta_s}  + E_{ns} \;\;.
\end{equation}
For a discussion see section IV of ref. \cite{myStructure}. 
In the analysis of our data, we assume that the finite size scaling
behavior of the two- and three-point functions is given by eq.~(\ref{finiteL}),
where, of course, the values of the constants depend on the quantity that
is considered.
Given the huge amount of data, we abstain from sophisticated fitting
with Ans\"atze motivated by eq.~(\ref{finiteL}).   Instead we consider pairs
of linear lattice sizes $L_1=L$, $L_2= 2 L$ and compute
\begin{equation}
\label{extrapol}
 G_{ex}(2 L) := G(2 L) + \frac{G(2 L) - G(L)}{2^{\Delta_s}-1} \;,
\end{equation}
where $G$ is the quantity under consideration. Eq.~(\ref{extrapol}) is
derived by inserting $L_1$ and $L_2$ into
eq.~(\ref{finiteL}) and solving the system of two equations with respect
to the non-singular (ns) part that remains in the limit $L \rightarrow \infty$.
As in the case of eq.~(\ref{ppt}), we 
use $\Delta_s=1.5113$ as numerical value for the scaling dimension.
In the analysis of the numerical data, for simplicity, we apply 
eq.~(\ref{extrapol}) to our estimates of the scaling dimension 
and the OPE coefficients $\lambda_{ijk}$ computed for finite lattice sizes
$L$. 
Note that the
first term on the right hand side of  eq.~(\ref{finiteL}) is subject to 
corrections to scaling. Given the small number of different linear 
lattice sizes $L$ that we consider here, these are not taken into account
in the extrapolation. Their effect is monitored by comparing the results
obtained by using different values of $L$ in the extrapolation.

\section{Numerical results}
\label{numerics}
\subsection{Preliminary simulations}
In order to check the $q$-dependence of our results for the $(q+1)$-state
clock model we have simulated 
the linear lattice size $L=120$ with stride $n_s=2$ for $q=8$, $16$, and
$32$. In all three cases $D=1.05$. In the case of $q=8$ we 
simulated at $\beta=0.56082390$, eq.~(\ref{betac105q8}) and for 
$q=16$ and $32$ at $\beta=0.56082418$, eq.~(\ref{betac105qL}).
The statistics is 139100, 139640, and 259820 measurements, respectively.
For each measurement, we performed $m=40$ measurements on the blocks.
For each block measurement we performed one sweep over the $3^3$ blocks.
We compared the results for the four different OPE coefficients for
all distances studied. At the level of our statistical accuracy, 
we find no dependence on $q$. Therefore, we are 
confident that the results obtained below for $q=32$ are essentially 
unaffected by the breaking of the $O(2)$-symmetry.

Furthermore we simulated the $O(2)$-symmetric $\phi^4$-model at 
$\lambda=2.1$ and $\beta=0.5091504$, eq.~(\ref{betacphi2.1}). 
We simulated the linear lattice size $L=120$ and 
used the stride $n_s=2$. We performed
81970 measurements with $m=60$ updates of the blocks for each measurement.
We compared our results for the OPE coefficients with those for the 
$(q+1)$-state clock model discussed above. In particular comparing 
with the $q=32$ case, we only find a difference that is clearly out 
of the error bars for $\lambda_{sss}$ at the distance $x=4$. In the
case of the $\phi^4$ model, we get $0.87927(40)$ and $0.84335(64)$ for the
directions $f$ and $d$, respectively. These numbers can be compared 
with $0.87698(30)$ and $0.84026(49)$ for the $(32+1)$-state clock model.
We also simulated the $\phi^4$-model for the stride $n_s=4$ and the 
linear lattice size $L=240$. Here we find no difference compared with the 
corresponding simulations of the $(32+1)$-state clock model discussed below.
Hence the small distance effects in the correlation functions are mainly 
caused by the lattice and the nearest neighbor interaction. 

Since the simulation of the $\phi^4$ model takes about three times
as much CPU time as that of the $(q+1)$-state clock model \cite{myClock}
and $16$ times as much memory is needed to store the configurations, we
simulated the $(q+1)$-state clock model with $q=32$ in the major part
of our study.

\subsection{Production runs using the $(32+1)$-state clock model}
In the major part of our study we simulated the $(32+1)$-state clock model
with the linear lattice sizes $L=240$, $480$, and $960$.  We performed 
measurements by using the strides $n_s=2$, $4$, and $6$.  In principle one
could do the measurements for these three different strides in the same
set of simulations. However, for simplicity, for a given simulation we 
performed measurements for one value of $n_s$ only.

Our final results are mainly based on the simulations with $n_s=6$. 
For $n_s=6$  we performed 70587, 11196, and 1272 update and measurement cycles
for $L=240$, $480$, and $960$, respectively.  On one core of an 
AMD EPYC 7351P 16-Core Processor the simulations using $n_s=6$ took about 
10 years in total.

\subsection{Scaling dimensions from the two-point correlation functions}
As a check  we extract the scaling dimensions $\Delta_{\phi}$, $\Delta_{s}$,
and $\Delta_{t}$  from the behavior of the two-point functions $g(x)$. 
In the first step we compute
\begin{equation}
\label{effDim}
 \Delta_{eff}(x,\Delta x) = - \frac{1}{2}  \frac{\ln(g(x+\Delta x)/g(x))}
 {\ln((x+\Delta x)/x)}  \;\;,
\end{equation}
where $\Delta x=n_s$, $\Delta x= \sqrt{2} n_s$, and $\Delta x= \sqrt{3} n_s$
for $r=a$, $f$ and $d$, respectively.

These results are extrapolated to the infinite volume by using
eq.~(\ref{extrapol}).

\begin{figure}
\begin{center}
\includegraphics[width=14.5cm]{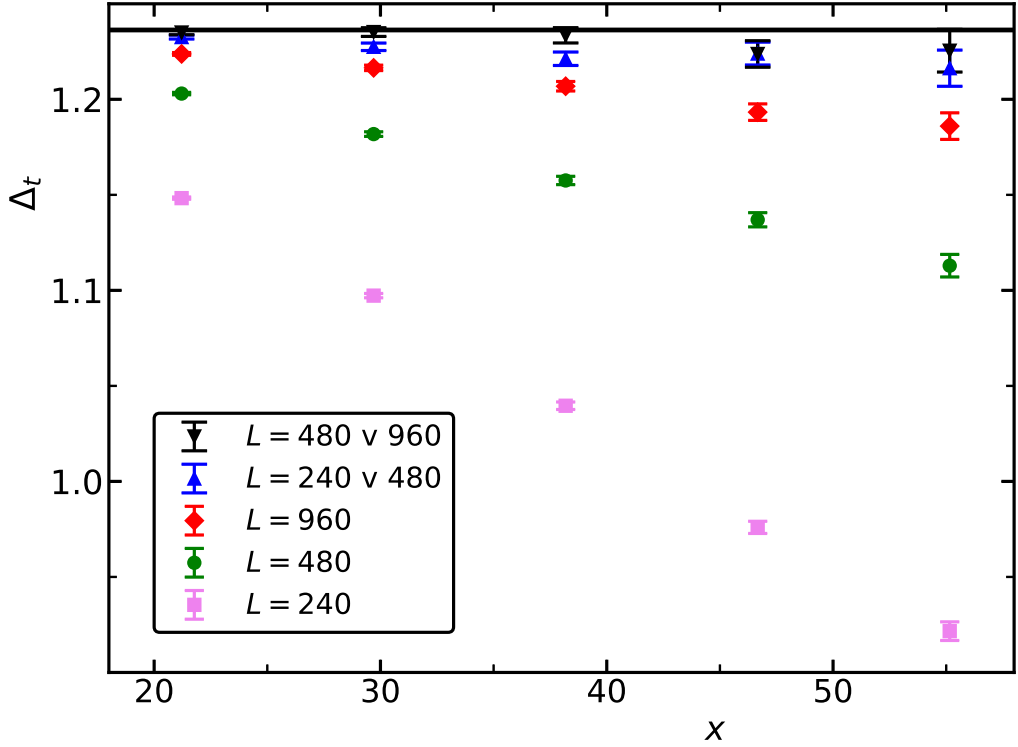}
\caption{\label{extraT}
We plot our numerical estimates of $\Delta_t$ as a function of the
distance $x$ between the lattice sites. Here we plot results for the 
direction $f$ only. The stride is $n_s=6$ throughout. We give estimates
computed for the linear lattice sizes $L=240$, $480$, and $960$ and
the extrapolations using the pairs $(240,480)$ and $(480,960)$ of
linear lattice sizes. For comparison we give the estimate obtained by using the 
conformal bootstrap method \cite{che19} as solid black line. 
}
\end{center}
\end{figure}

In Fig. \ref{extraT}  we demonstrate the effectiveness of the extrapolation.
We give the results for $\Delta_t$ obtained for the linear lattice sizes 
$L=240$, $480$, and $960$.  The measurements are performed with the stride 
$n_s=6$. Here we give results for the direction $f$ only. We see a clear
dependence of the results on $L$ that increases with increasing distance.
In contrast, the extrapolated results for $(L_1,L_2)=(240,480)$ and
$(L_1,L_2)=(480,960)$  differ only by little.
Note that the error bars given in Fig. \ref{extraT}, as for the figures
below, are purely
statistical, indicating one standard deviation. Furthermore, for
a given $n_s$, the results for different distances are obtained from the
same simulations. Hence there is a statistical cross-correlation.

\begin{figure}
\begin{center}
\includegraphics[width=14.5cm]{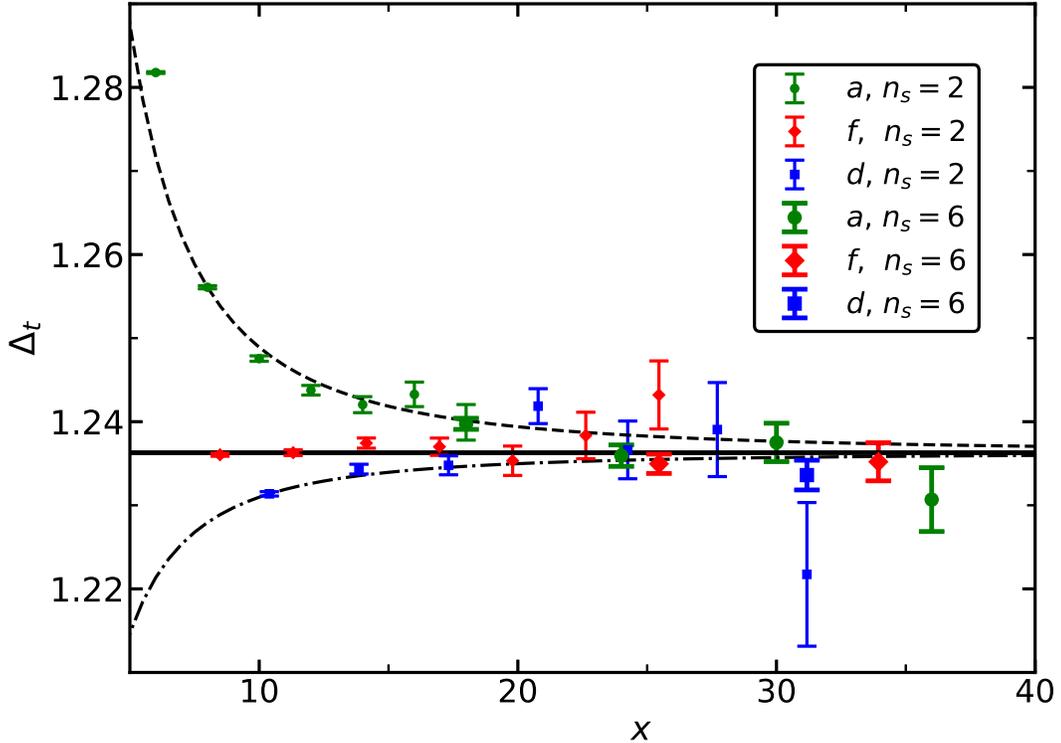}
\caption{\label{Talldir}
We plot $\Delta_t$ obtained by extrapolating our results for 
$(L_1,L_2)=(480,960)$ as a function of the
distance $x$ between the lattice sites. These results are obtained 
by using the strides $n_s=2$ and $6$. We omit the results for $n_s=4$
to keep the figure readable. $a$, $f$, and $d$ denote the three different 
directions that we consider.
For comparison we give the estimate obtained by using the
conformal bootstrap method \cite{che19} as solid black line. 
In addition we give dashed and dash-dotted lines that include 
an estimate of corrections to scaling for the directions $a$ and $d$, 
respectively. For a discussion see the text.
}
\end{center}
\end{figure}

Next we check for the effect of operators with higher dimension in 
the same channel. The effect should decay with increasing distance 
between the two sites.  In Fig. \ref{Talldir} we plot extrapolated results for 
$(L_1,L_2)=(480,960)$ of $\Delta_t$. Data are taken from our runs for the 
strides $n_s=2$ and $6$.  We give results for all three directions 
that we consider.  
Similar to the case of the Blume-Capel model on the simple cubic 
lattice we find that the amplitude of corrections is quite different 
for different directions \cite{myStructure}. 
For the direction $f$ the deviation at small distances
$x$ are the smallest, while for $d$ they are the largest.  To 
check whether it is plausible that corrections due to the violation 
of rotational symmetry by the lattice dominate, as discussed in 
section \ref{observables} above, we plot
$D + c x^{-2.02}$ as dashed and dash-dotted lines for the directions
$a$ and $d$, respectively. For $D$ we take the value of $\Delta_t$ obtained 
by the CB method. The coefficient $c$ is simply chosen such that the numerical
estimate of $\Delta_t$ at $x=8$ and $6 \times \sqrt{3}$ for the directions 
$a$ and $d$ are matched, respectively.

The observations are similar for $\Delta_s$ and $\Delta_{\phi}$ that 
we do not discuss in detail here.

Our numerical result for $\Delta_t$ is consistent with that obtained
by using the CB method \cite{che19} and previous Monte Carlo simulations. 
However, we do not reach the accuracy of \cite{che19} and the lattice result
\cite{HaVi11}. As our final estimate, we might quote
$\Delta_t= 1.2352(23)$  from the extrapolation of $(L_1,L_2)=(480,960)$ for
the direction $f$ and the pair of distances $(12,18) \times \sqrt{2}$.
Looking at Figs. \ref{extraT} and \ref{Talldir} it seems plausible that
for this choice, systematical errors due to the finite value of $x$ and 
due to the imperfection of the extrapolation to the infinite volume are
not larger than the statistical error.

In a similar way we get $\Delta_{\phi}=0.51953(32)$ from 
the extrapolation of $(L_1,L_2)=(480,960)$, the stride $n_s=2$, the
direction $f$, and the pair of distances $(12,14)  \times \sqrt{2}$,
or $\Delta_{\phi}=0.51864(52)$ for
the stride $n_s=6$ and the pair of distances $(12,18)  \times \sqrt{2}$.
Finally we obtain from the measurements with the stride $n_s=6$,  the
direction $f$, and the  pair of distances $(12,18)  \times \sqrt{2}$
the estimate $\Delta_s= 1.5098(21)$.

\subsection{The OPE coefficients and the three-point functions}
Here we follow a similar procedure as for the scaling dimensions.
In the first step, we compute estimates of $\lambda_{ijk}$ for 
given linear lattice sizes $L$ by using eq.~(\ref{ppt}) and analogous 
equations. Then we extrapolate to the infinite volume by using 
eq.~(\ref{extrapol}). 
Similar to ref. \cite{myStructure} we have measured the three-point
function for two different geometries that we denote by $f$ and $d$. 
It turns out that small distance corrections are smaller for $f$.
Final results are however fully consistent. Therefore in the following 
we restrict the detailed discussion on geometry $f$.

\begin{figure}
\begin{center}
\includegraphics[width=14.5cm]{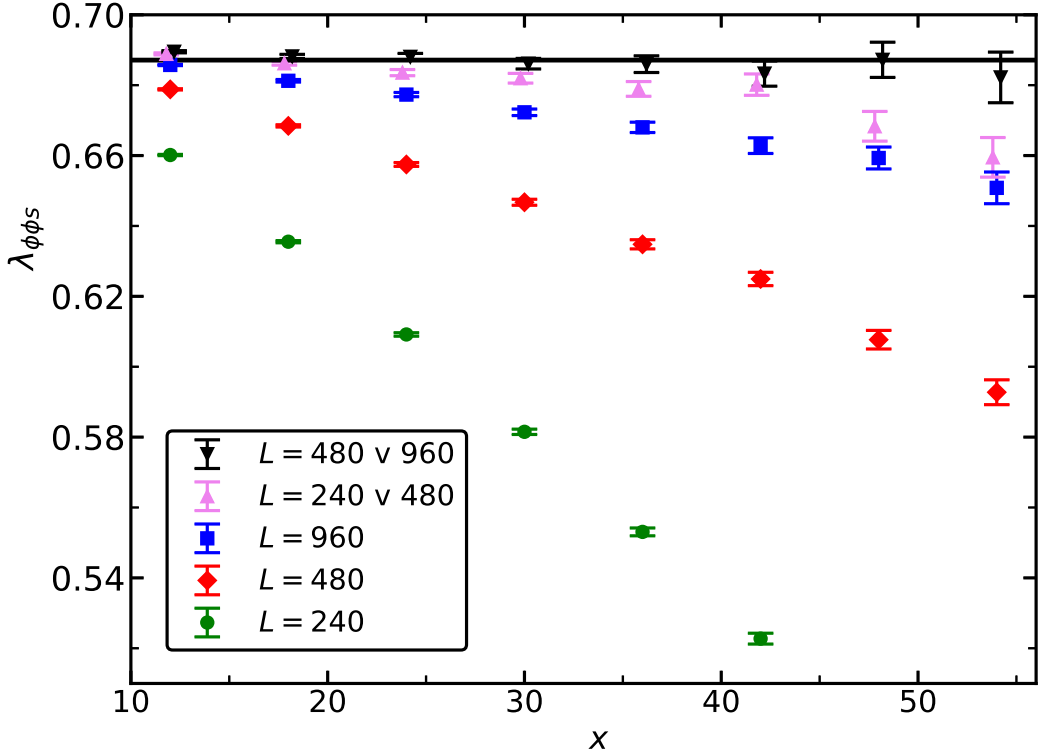}
\caption{\label{structure1}
We plot our numerical results for $\lambda_{\phi \phi s}$ as a function of the 
distance $x$. Here we consider simulations with stride
$n_s=6$ and three-point functions for the geometry $f$. We give results for
the linear lattice sizes $L=240$, $480$ and $960$. These results are
extrapolated by using eq.~(\ref{extrapol}) for the pairs $(240,480)$ 
and $(480,960)$ of linear lattice sizes.
The distance $x$ for the pair $(240,480)$ is slightly shifted to make the
figure more readable.
For comparison we give the estimate obtained by using the
conformal bootstrap method \cite{che19} as solid black line. 
}
\end{center}
\end{figure}

In Fig. \ref{structure1} we plot results for $\lambda_{\phi \phi s}$
obtained from simulations with the stride $n_s=6$ and the linear
lattice sizes $L=240$, $480$, and $960$. In addition we give the 
results of the extrapolation using eq.~(\ref{extrapol}) and the 
pairs of lattice sizes $(L_1,L_2)=(240,480)$ and $(L_1,L_2)=(480,960)$.
We see a clear 
dependence of the results on $L$ that increases with increasing distance $x$.
In contrast, the extrapolated results for $(L_1,L_2)=(240,480)$
and $(L_1,L_2)=(480,960)$  differ only by little. Only for distances
$x \ge 48$ the estimate obtained from the pair $(L_1,L_2)=(240,480)$ decreases
significantly with increasing distance $x$. Based on this observation, we conclude that  
the extrapolation for $(L_1,L_2)=(480,960)$ is reliable in the range of distances 
$x$ considered below. We have checked that the same also holds for the other 
three OPE coefficients that we study.

\begin{figure}
\begin{center}
\includegraphics[width=14.5cm]{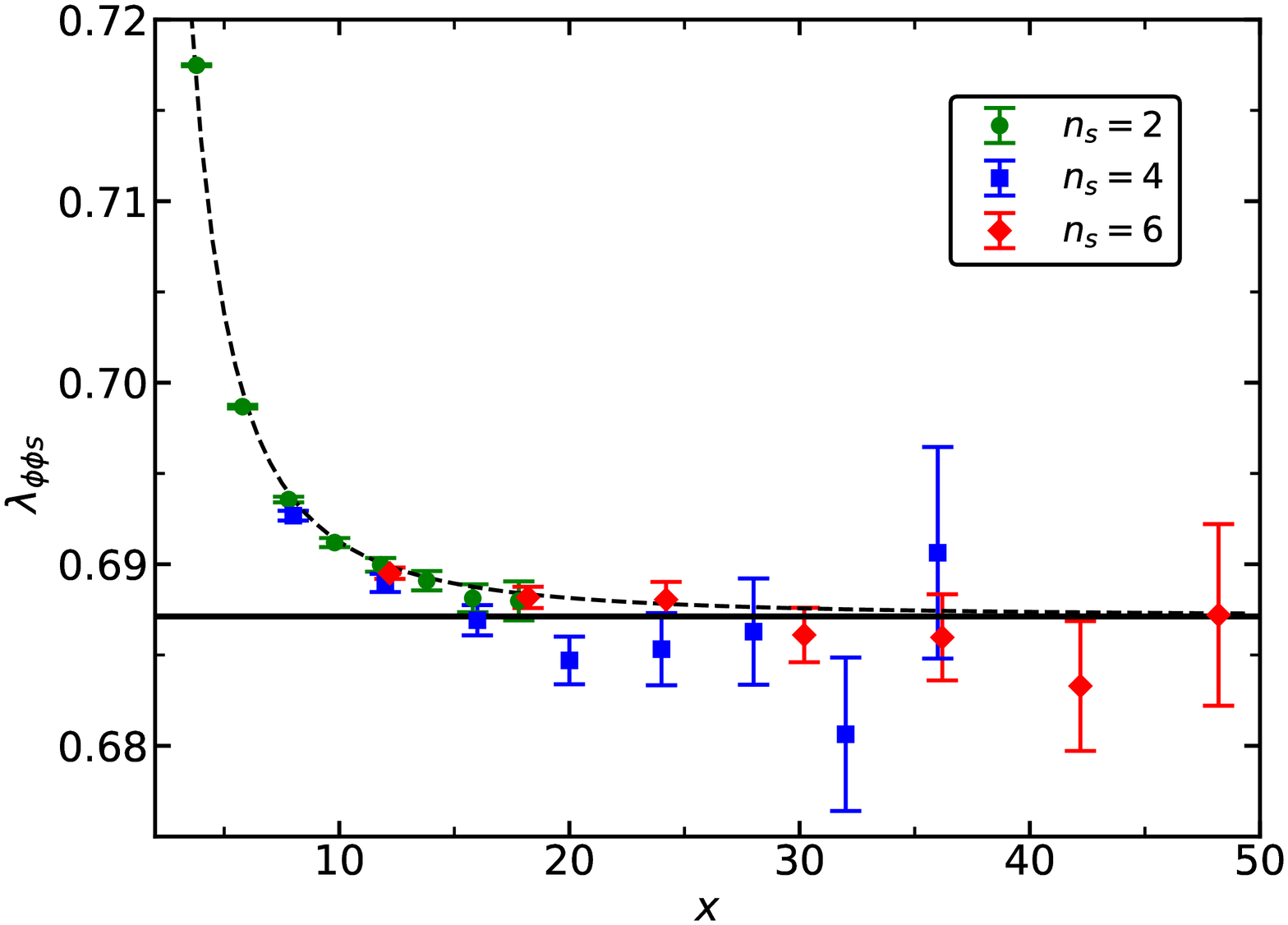}
\caption{\label{structure2}
We plot our numerical results for $\lambda_{\phi \phi s}$ as a function of the
distance between the lattice sites. Here we consider simulations with stride
$n_s=2$, $4$, $6$ and three-point functions for the geometry $f$.
We give results for the extrapolation of the lattice sizes
$(L_1,L_2)=(480,960)$. The values of $x$ for $n_s=2$ and $6$ 
are slightly shifted to reduce the overlap of the symbols.
For comparison we give the estimate obtained by using the
conformal bootstrap method \cite{che19} as solid black line.
The dashed line contains in addition a correction $\propto x^{-2.02}$.
}
\end{center}
\end{figure}

Next in Fig. \ref{structure2} we plot the extrapolated results for 
$(L_1,L_2)=(480,960)$ obtained for the strides $n_s=2$, $4$ and $6$.
  In order to
check whether it is plausible that corrections due to the violation
of rotational symmetry by the lattice dominate, we plot
$l + c x^{-2.02}$ as dashed line, where $l$ is the estimate of 
$\lambda_{\phi \phi s}$ obtained by the CB method and $c$ is chosen such 
that our numerical estimate for the distance $x=6$ is matched. Indeed,
the data fall reasonably well on the dashed line. In the case of $n_s=4$
and $x=20$ there is a deviation by about $2.6$ standard deviations.
The deviations at $x=12$ and $16$ go in the same direction. Since the estimates
at different distances are obtained from the same simulations, there is a 
statistical correlation between them. Hence it is still reasonable to 
attribute these deviations to statistical fluctuations.
For the stride $n_s=6$ we get $\lambda_{\phi \phi s}=0.6881(10)$ at
$x=24$. The dashed line suggests that for $x=24$ the finite $x$ effect is
smaller than the statistical error. Hence one might base a final result
on this estimate.

\begin{figure}
\begin{center}
\includegraphics[width=14.5cm]{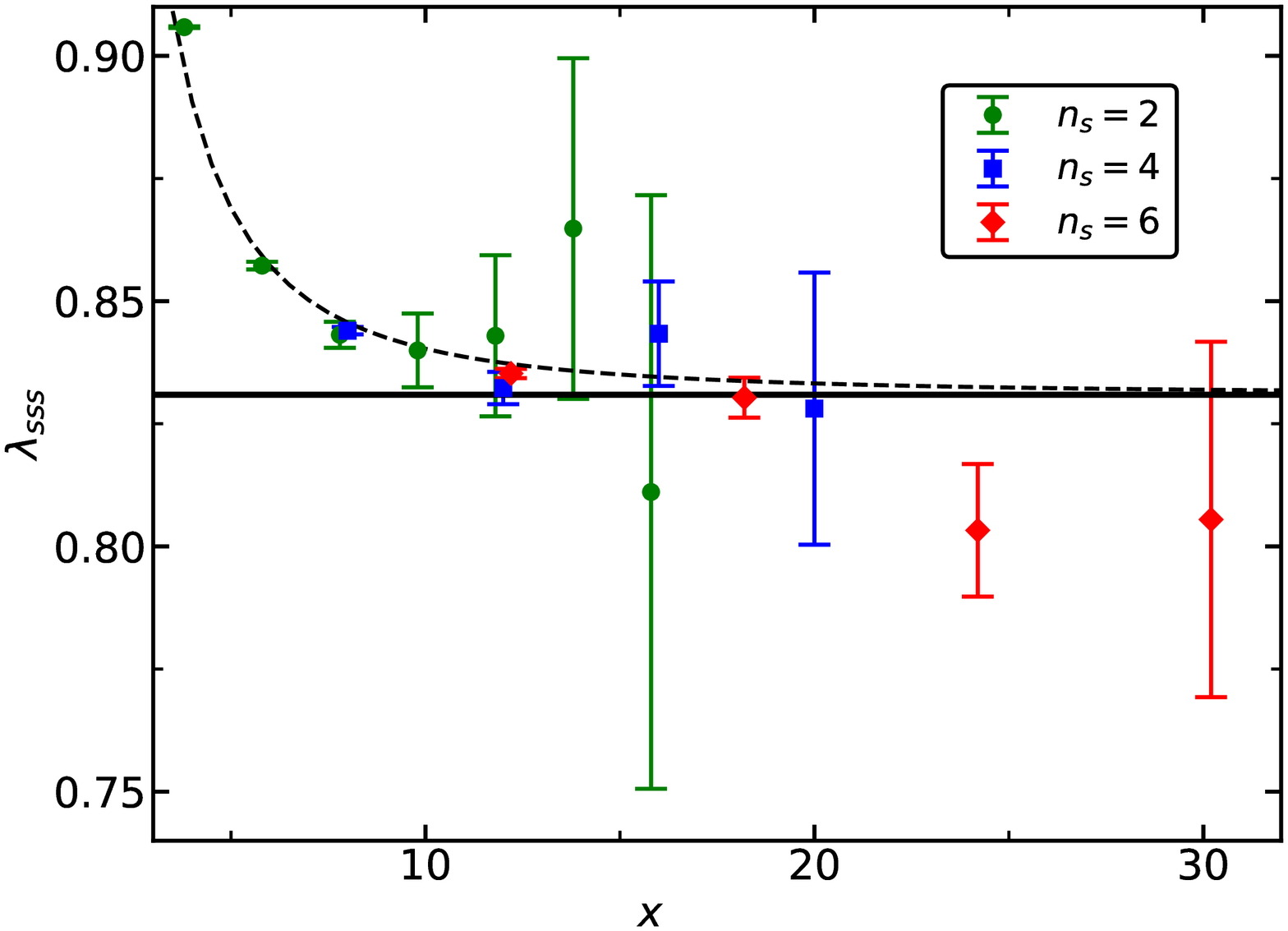}
\caption{\label{structure3}
We plot our numerical results for $\lambda_{s s s}$ as a function of the
distance between the lattice sites. Here we consider simulations with stride
$n_s=2$, $4$, $6$ and three-point functions for the geometry $f$. 
We give results for the extrapolation of the lattice sizes 
$L=480$ and $960$. 
The values of $x$ for $n_s=2$ and $6$
are slightly shifted to reduce the overlap of the symbols.
For comparison we give the estimate obtained by using the
conformal bootstrap method \cite{che19} as solid black line.
The dashed line contains in addition a correction $\propto x^{-2.02}$.
}
\end{center}
\end{figure}

Next let us discuss the numerical results for $\lambda_{sss}$. First we
convinced ourself that also here the extrapolation by using 
eq.~(\ref{extrapol}) is effective. In Fig. \ref{structure3} we give results
of the extrapolation using the linear lattice sizes $L=480$ and $960$
for the strides $n_s=2$, $4$, and $6$. 
The relative statistical error is larger than for  $\lambda_{\phi \phi s}$.
The effect of the variance reduction is more important than for 
$\lambda_{\phi \phi s}$. Going to larger distances, it is mandatory to use
larger block sizes.

Similar to Fig. \ref{structure2}, we plot
$l + c x^{-2.02}$ as dashed line, where $l$ is the estimate of
$\lambda_{sss}$ obtained by the CB method and $c$ is chosen such
that our numerical estimate for the distance $x=6$ is matched.
Based on that it seems plausible that for the stride $n_s=6$ at 
the distance $x=18$, the small distance error is at most of similar size
as the statistical one. We read off $\lambda_{s s s}=0.8303(41)$.

\begin{figure}
\begin{center}
\includegraphics[width=14.5cm]{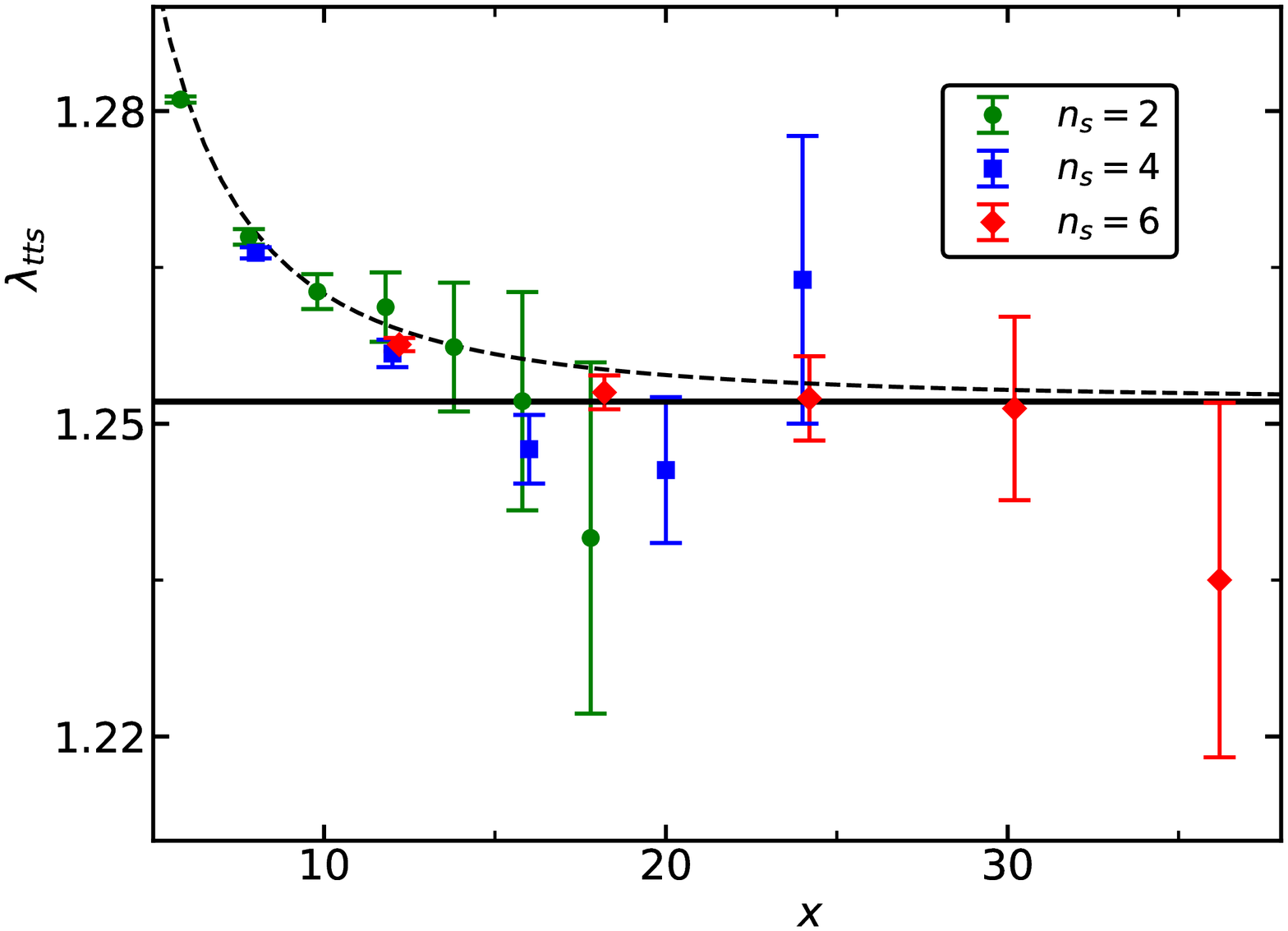}
\caption{\label{structure4}
We plot our numerical results for $\lambda_{t t s}$ as a function of the
distance $x$ between the lattice sites. Here we consider simulations with stride
$n_s=2$, $4$, $6$ and three-point functions for the geometry $f$.
We give results for the extrapolation of the lattice sizes
$L=480$ and $960$. 
The values of $x$ for $n_s=2$ and $6$
are slightly shifted to reduce the overlap of the symbols.
For comparison we give the estimate obtained by using the
conformal bootstrap method \cite{che19} as solid black line.
The dashed line contains in addition a correction $\propto x^{-2.02}$.
The values of $x$ for $n_s=2$ and $6$
are slightly shifted to reduce the overlap of the symbols.
}
\end{center}
\end{figure}

Next, in Fig. \ref{structure4} we plot our extrapolated results for 
$\lambda_{t t s}$ obtained from the simulations with the strides
$n_s=2$, $4$, and $6$. 
Similar to Fig. \ref{structure2} we plot
$l + c x^{-2.02}$ as dashed line, where $l$ is the estimate of
$\lambda_{tts}$ obtained by the CB method and $c$ is chosen such
that our numerical estimate for the distance $x=6$ is matched.
The  final result could be based on the estimate 
$\lambda_{t t s}=1.2530(16)$
obtained by using the stride $n_s=6$ at the distance $x=18$.

\begin{figure}
\begin{center}
\includegraphics[width=14.5cm]{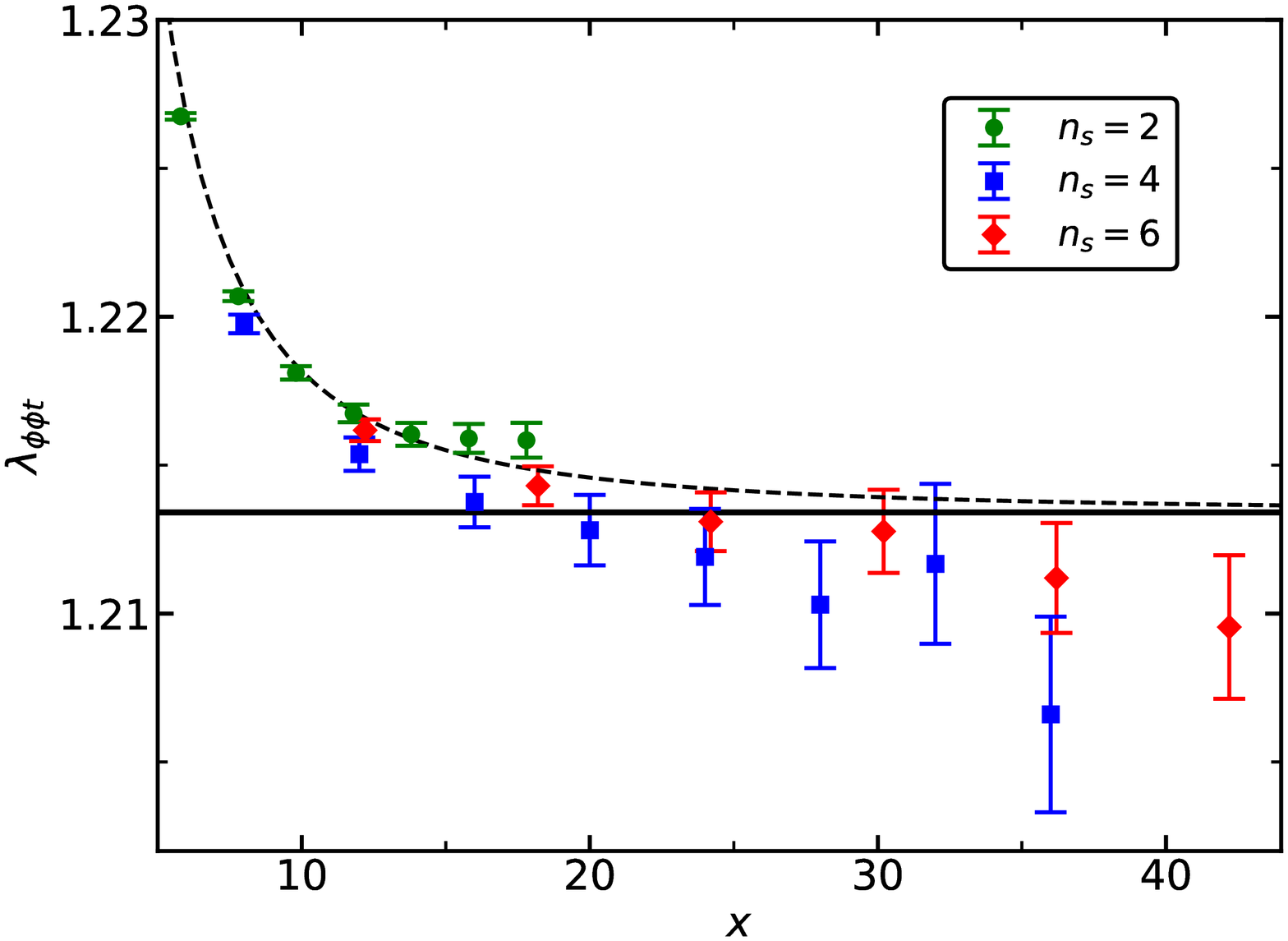}
\caption{\label{structure5}
We plot our numerical results for $\lambda_{\phi \phi t}$ as a function of the
distance between the lattice sites. Here we consider simulations with stride
$n_s=2$, $4$, $6$ and three-point functions for the geometry $f$.
We give results for the extrapolation of the lattice sizes
$L=480$ and $960$.
The values of $x$ for $n_s=2$ and $6$
are slightly shifted to reduce the overlap of the symbols.
For comparison we give the estimate obtained by using the
conformal bootstrap method \cite{che19} as solid black line.
The dashed line contains in addition a correction $\propto x^{-2.02}$. 
Note that we have multiplied our numbers, which are based on 
eqs.~(\ref{gpp},\ref{gtt},\ref{Gppt}),
by a factor of $\sqrt{2}$ to match with the conventions of ref. \cite{che19}.
}
\end{center}
\end{figure}

Finally, in Fig. \ref{structure5} we plot our numerical results for 
$\lambda_{\phi \phi t}$.   
Similar to Fig. \ref{structure2} we plot
$l + c x^{-2.02}$ as dashed line, where $l$ is the estimate of
$\lambda_{\phi \phi t}$ obtained by the CB method and $c$ is chosen such
that our numerical estimate for the distance $x=6$ is matched.

We read off $\lambda_{\phi \phi t}=1.214(7)$
for $x=18$ and the stride $n_s=6$. For the distance $x=24$ we get 
$\lambda_{\phi \phi t}=1.213(10)$ instead.
Note that we have multiplied our numbers, which are based on
eqs.~(\ref{gpp},\ref{gtt},\ref{Gppt}),
by a factor of $\sqrt{2}$ to match with the conventions of ref. \cite{che19}.

\section{Summary and discussion}
\label{summary}
We have demonstrated that OPE coefficients for the three-dimensional
XY universality
class can be determined by using Monte Carlo simulations of a lattice model
with a relative error of about $1 \%$ or less. To this end,
we have simulated the improved $(32+1)$-state clock model using 
linear lattice sizes up to $L=960$. The outline of the study is 
similar to that of ref. \cite{myStructure}, where we studied the 
Ising universality class. The key ideas are variance reduced 
estimators of the two- and three-point correlation function and an 
extrapolation  to the infinite volume.

Our results are fully consistent with those recently obtained by using the 
conformal bootstrap (CB) method \cite{che19}, further confirming that the
CB method and the lattice model examine the same RG fixed point. One has
to note however that the estimates obtained by using the CB method are by
about two orders of magnitude more precise than those obtained here.

There is still room for improvement. For example,
the measurement, which takes considerably more CPU time than the generation
of the configurations, could be easily parallelized and could hence be speeded
up for example by running it on graphics processing units (GPUs). 


\section{Acknowledgement}
This work was supported by the Deutsche Forschungsgemeinschaft (DFG) under 
the grant No HA 3150/5-1.

\end{document}